\documentclass[%
 reprint,
 showpacs,preprintnumbers,
 amsmath,amssymb,
 aps,
prl,
]{revtex4-1}
\usepackage{dcolumn}
\usepackage{subfigure}
\usepackage{graphicx,color}
\usepackage{mathrsfs}

\makeatletter
\def\btt#1{\texttt{\@backslashchar#1}}
\DeclareRobustCommand\bblash{\btt{\@backslashchar}} \makeatother

\begin{document}

\title{Stable $2R$ van der Waals heterostructures of NbS$_2$ and $M$Se$_2$ for $M$=Mo and W}

\author{Tetsuro Habe}
\affiliation{Nagamori Institute of Actuators, Kyoto University of Advanced Science, Kyoto 615-8577, Japan}

\date{\today}

\begin{abstract}
In this letter, we investigate the stable and commensurate van der Waals heterostructures of metallic and semiconducting $1H$ transition-metal dichalcogenides, NbS$_2$ and MoSe$_2$ (WSe$_2$), which possess almost the same lattice constant of the pristine honeycomb structure.
In the most stable structure, the metallic and semiconducting layers are stacked in a similar manner to $3R$ stacking but the period is a pair of a metallic layer and a semiconducting layer.
The heterostructure aligns the spin-polarization in each valley among all layers and induces spin-selective charge transfer between the metallic and semiconducting layers.
Especially in hetero-trilayers, the electronic spin is conserved due to mirror symmetry along the out-of-plane axis in contrast to the $3R$ stacking structure.
A drastic enhancement of spin Hall effect is numerically shown as an example of electronic spin transport phenomena in the hetero-trilayers.
\end{abstract}

\maketitle
%\section{Introduction}
Stacking structures of atomic layered materials drastically change the electronic properties and have attracted much attention recently.
The atomic layered materials consist of atomically thin layer-like crystals stacked in the vertical direction to the layers.
Since the layers are bonded weakly to each other by van der Waals (vdW) interaction, a single layer can be cleaved.
The experimental method enables us to produce so-called vdW heterostructure by stacking layers cleaved from different atomic layered materials\cite{Geim2013} and to study fascinating electronic properties, e.g., phase transitions in twisted bilayer graphene with a moir\'{e} pattern.\cite{Bistritzer2011-1,Bistritzer2011-2}
At this time, other experimental methods are also available for producing the vdW heterostructures.
Chemical vapor deposition and molecular-beam-epitaxy are available to realize a stable vdW heterostructure with the lowest electronic energy.\cite{Lee2012,Chhowalla2013,Bergeron2017,Zhan2017,Kazzi2018,Maxime2018}
In this letter, we consider the stable vdW heterostructures of $1H$ transition-metal dichalcogenides (TMDCs) with the almost same lattice constant and reveal fascinating electronic properties due to the electronic structure.
We performed first-principles calculation for the total electronic energy and showed the stacking structure of NbS$_2$ and MoSe$_2$ (WSe$_2$) to be a new structure, 2$R$. 
Moreover, numerical calculations in linear response theory reveal that the novel stacking drastically enhances the spin Hall conductivity.

%\section{Stable stacking structure}
\begin{figure}[htbp]
\begin{center}
 \includegraphics[width=80mm]{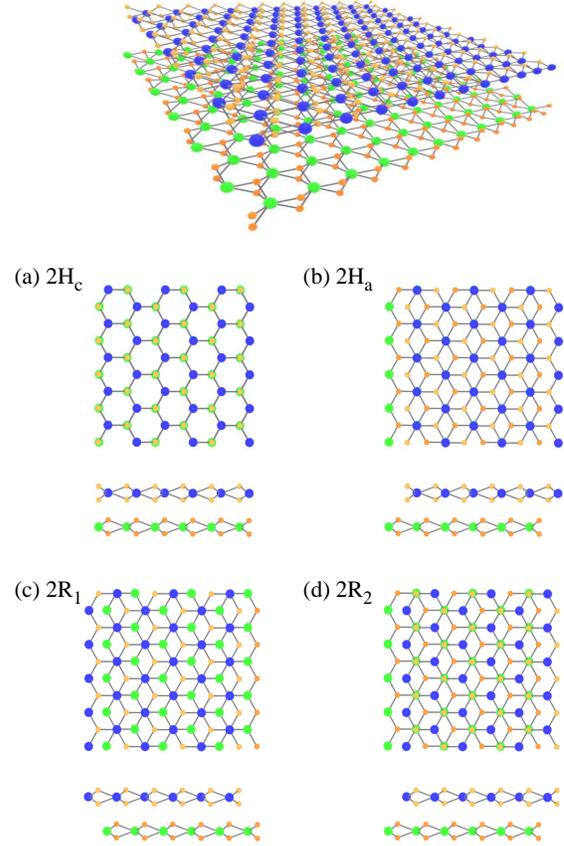}
\caption{The schematics of van der Waals bilayer of two $1H$ TMDCs. In (a)-(d), the top and horizontal views are presented for four high-symmetric stacking structures, $2H_c$, $2H_a$, $2R_1$, and $2R_2$.
 }\label{fig_schematics}
\end{center}
\end{figure}
The stable stacking structure of TMDCs depends on the atomic species of each layer.
In the cases of the pristine atomic layered materials or heterostructures of semiconducting TMDCs, $2H_a$ or $2H_c$ stacking is stable in terms of the total electronic energy.\cite{He2014,Debbichi2014,Sharma2014}
In semiconducting TMDCs, MoSe$_2$ and WSe$_2$, layers are stacked in the $2H_c$ stacking structure shown in Fig. \ref{fig_schematics}(a).
On the other hand, NbS$_2$, a metallic TMDC, consists of $1H$ monolayers but it is in $2H_a$ stacking in Fig. \ \ref{fig_schematics}(b).
In these two structures, adjacent layers possess spatially inverted honeycomb structures but the relative shift of these layers is different.
Although electronic states in $1H$ crystal possess spin-valley locking due to inversion symmetry breaking,\cite{Xiao2012} these $2H$-type stacking structures reduce or eliminate the spin-valley correlation in the multilayer crystals.
Since spatial inversion exchanges electronic states between the $K$ and $K'$ valleys, both spin states are degenerated in the same valley of $2H_a$ or $2H_c$ bilayer crystals.
Two bilayers in $2R_1$ and $2R_2$ correspond to two types of $3R$ stacking as shown in Figs.\ \ref{fig_schematics}(c) and (d), respectively.
Although the spin-valley locking is preserved in $2R$ stacking structures,\cite{Kormanyos2018} the stacking structure is not the most stable structure of pristine TMDCs.
Thus, the pristine multilayer TMDC cannot preserve the fascinating property, the spin-valley locking, in the most stable structure.

\begin{figure}[htbp]
\begin{center}
 \includegraphics[width=80mm]{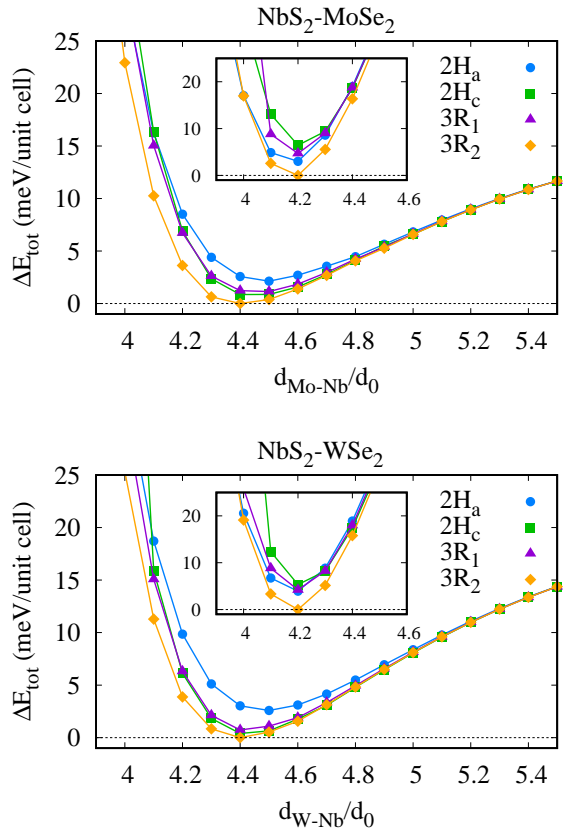}
\caption{The variation of total electronic energy with the inter-layer distance. The electronic energy is given with respect to the lowest one and the inter-layer distance is normalized by the distance between the Nb-sublayer and S-sublayer in NbS$_2$. Four symbols represent different stacking structures. The insets present the numerical results spinless but including the effect of van der Waals interaction.
 }\label{fig_energy_of_vdW-hetero}
\end{center}
\end{figure}
In the vdW heterostructures of the metallic and semiconducting TMDCs, it is theoretically shown that a different stacking from that in the pristine TMDCs appears as the most stable structure with preserving the commensurate honeycomb lattice.
Generally speaking, atomic layered materials possess different lattice constants due to the atomic species and thus the vdW heterostructure of such layers forms an incommensurate structure, i.e., moir\'e structure, and loses the original periodicity.
However, the previous paper shows that the lattice constant of honeycomb structure is almost the same, $a\simeq 3.320$\AA, among pristine NbS$_2$, MoSe$_2$, and WSe$_2$\cite{Habe2022-1}.
Thus, these layers form commensurate vdW hetero-multilayers, i.e., crystalline vdW heterostructures.
Since the true grand state gives the smallest total energy of the electronic system, it can be determined with performing the first-principles calculations of the electronic structure for different stacking structures.
The numerical calculation is performed with using Quantum-espresso,\cite{quantum-espresso} a package of numerical codes for first-principles calculations in density functional theory (DFT), with projector augmented wave method including spin-orbit coupling within generalized gradient approximation (GGA)\cite{Perdew1996}.
In the numerical calculations, Perdew-Burke-Emzerhof exchange-correlation functional is adopted as the GGA functional.
The energy cut-off is 50Ry for the plane wave basis and 500Ry for the charge density on $12\times12\times1$ mesh in the first Brillouin zone given by the primitive reciprocal vectors.
The convergence criterion 10$^{-8}$ Ry is adopted for the self-consistent field calculation.
The schematics of bilayer crystals are given in Fig.\ \ref{fig_schematics}, where the bottom layer is the semiconducting one, i.e., MoSe$_2$ or WSe$_2$, and the top layer is NbS$_2$.
In Fig.\ \ref{fig_energy_of_vdW-hetero}, the total electronic energy is presented for different staking of the crystalline vdW hetero-bilayers with different inter-layer distances.
Here, the inter-layer distance is defined by that between the Nb-sublayer and the Mo(W)-sublayer in the unit distance $d_0$ between the Nb-sublayer and the S-sublayer of NbS$_2$.
The lattice constant of the honeycomb structure and the atomic positions in the perpendicular direction to the layers are optimized for each inter-layer distance. 
Here, the crystal structure optimization is performed with using vc-relax code of Quantum-espresso with the convergence threshold for forces 10$^{-4}$ Ry/Bohr and that for stresses 10$^{-1}$ kbar.
The insets also present the total electronic energy calculated by DFT calculations spinless but including van der Waals interaction with vdW-DF2.\cite{Lee2010}
The numerical results clearly show that the $2R_2$ stacking structure is the most stable one for both hetero-bilayers.
In the following calculations, the inter-layer distance obtained by spinful calculations is adopted for the investigation of spin properties.

The stability of $2R_2$ stacking is consistent with the difference of the pristine NbS$_2$ and MoSe$_2$ (WSe$_2$).
The semiconducting TMDCs prefer the $2H_c$ stacking structure, where all transition-metal atoms, Mo or W, are aligned just below or above the chalcogen atoms in the adjacent layer.
In the $2H_a$-stacked NbS$_2$, Nb atoms avoid being close to the chalcogen atoms in the adjacent layer.
Similarly, in the $2R_2$ stacking structure, Mo or W minimizes the distance from chalcogen atoms and Nb maximizes that.
Therefore, the $2R_2$ structure is attributed to the difference in the relation between the transition-atom and the chalcogen atom in NbS$_2$ and MoSe$_2$ (WSe$_2$).

\begin{figure}[htbp]
\begin{center}
 \includegraphics[width=80mm]{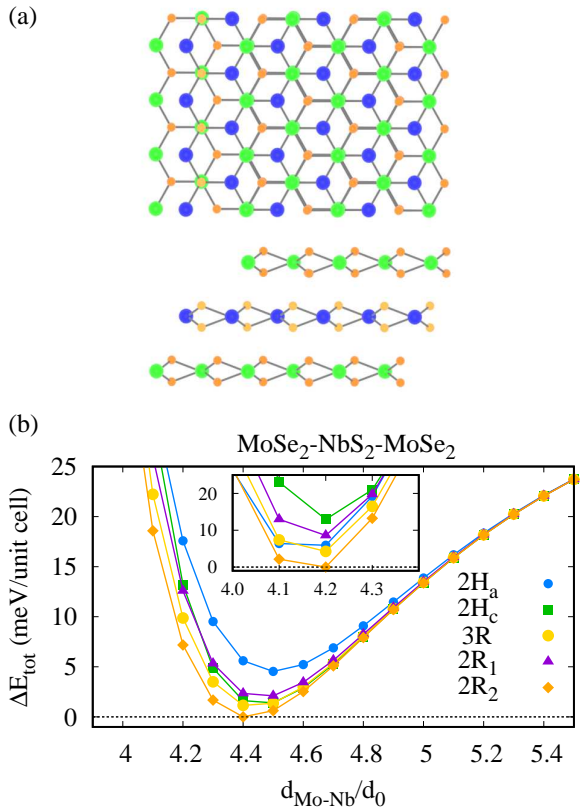}
\caption{The variation of total electronic energy with the inter-layer distance for vdW hetero-trilayer MoSe$_2$-NbS$_2$-MoSe$_2$. }\label{fig_energy_of_vdW-trilayer}
\end{center}
\end{figure}
In the hetero-trilayer consisting of one metallic layer sandwiched by two semiconducting layers, $2R_2$ stacking is the most stable structure in comparison with $2H_a$, $2H_c$, $3R$, and $2R_1$.
Here, the $2R_\mu$ trilayer consists of the $2R_\mu$ bilayer covered by another semiconducting layer equally aligned with the bottom layer in Figs.\ \ref{fig_schematics}(c) and \ref{fig_schematics}(d).
Obviously, in the $2R_2$ trilayer, Nb atoms maximize the distance from chalcogen atoms and Mo or W atoms minimize that among the three layers.
In Fig.\ \ref{fig_energy_of_vdW-trilayer}, the variation of total electronic energy for each type of MoSe$_2$-NbS$_2$-MoSe$_2$ hetero-trilayer is presented with the interlayer distance.
The numerical calculation shows that $2R_2$ stacking minimizes the total electronic energy.
In the case of WSe$_2$-NbS$_2$-WSe$_2$, $2R_2$ stacking is also realized as the most stable structure.

\begin{figure}[htbp]
\begin{center}
 \includegraphics[width=80mm]{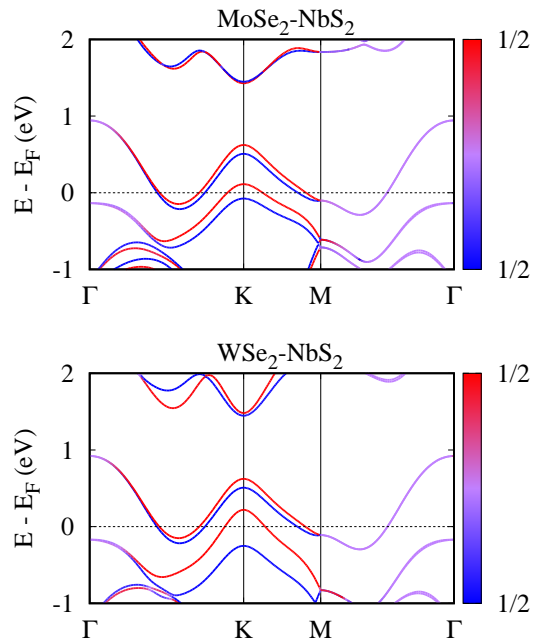}
\caption{The band structures of vdW bilayer MoSe$_2$-NbS$_2$ and WSe$_2$-NbS$_2$ with the spin component.
 }\label{fig_band-structure_vdW-bilayer}
\end{center}
\end{figure}
The $2R_2$ vdW heterostructure enhances the spin polarization in the $K$ and $K'$ valleys without any extrinsic fabrication, e.g., magnetic ordering and charge doping.
In Fig.\ \ref{fig_band-structure_vdW-bilayer}, the band structure and the spin component of electronic states are presented for $2R_2$ vdW hetero-bilayer MoSe$_2$-NbS$_2$ and WSe$_2$-NbS$_2$ with the spin axis normal to the layers.
Here, the first-principles band structures are calculated with using Quantum-espresso and the electronic spinful states are obtained with using Wannier90,\cite{wannier90} a numerical package for calculating the hopping parameters between the maximally localized Wannier functions from a first-principles band structure.
In the $K$ valley, the two spin states are well separated due to the intrinsic large spin split in each monolayer though mirror symmetry is broken in the out-of-plane axis.
Around the Fermi energy, there are two spin-split bands attributed to the NbS$_2$ and MoSe$_2$ (WSe$_2$) layers.
The upper band mainly consists of the Wannier orbitals in the NbS$_2$ layer and the lower one is constructed by those in the semiconducting layer.
In both of the spin-split bands, the up-spin states possess a higher energy in the $K$ valley because of the stacking structure.
Both up-spin and down-spin states of the upper band can be found in the Fermi energy but the lower one provides only the up-spin states in the $K$ valley.
Since the lower band is attributed to that in the semiconducting $1H$-TMDC layer, it is shown that holes are transferred only to the up-spin states in the semiconducting layer from the NbS$_2$ layer around the $K$ point.
On the other hand, no charge transfer is observed around the $\Gamma$ point. 
In the $K'$ valley, the down-spin states in the lower band attract holes from the NbS$_2$ layer due to time-reversal symmetry.
Therefore, the $2R_2$ stacking structure provides spin-selective charge transfer in the $K$ and $K'$ valleys but the hetero-bilayer cannot preserve two spins rigorously due to mirror symmetry breaking along the out-of-plane direction.

\begin{figure}[htbp]
\begin{center}
 \includegraphics[width=80mm]{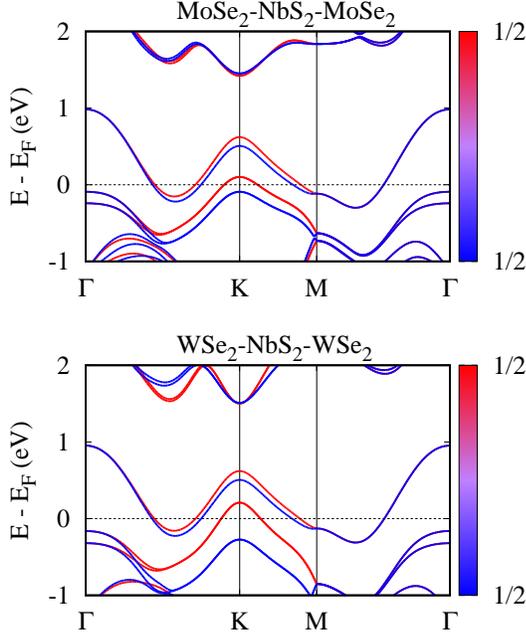}
\caption{The band structures of vdW trilayer MoSe$_2$-NbS$_2$-MoSe$_2$ and WSe$_2$-NbS$_2$-WSe$_2$ with the spin component.
 }\label{fig_band-structure_vdW-trilayer}
\end{center}
\end{figure}
The $2R_2$ hetero-trilayers also induce spin-selective charge transfer in the two valleys, and they can conserve two spin states as well.
In Fig.\ \ref{fig_band-structure_vdW-trilayer}, the electronic band structures of the $2R_2$ trilayers are presented with the spin component of each electronic state.
Around the $K$ point, the band structure is quite similar to that of the $2R_2$ hetero-bilayer shown in Fig.\ \ref{fig_band-structure_vdW-bilayer}.
However, the lower spin-split band around the Fermi energy is doubly degenerated because of two semiconducting layers sandwiching the NbS$_2$ layer.
In fact, the band is split around the $\Gamma$ point in contrast to the case of hetero-bilayer.
Around the Fermi energy, the two spin states in these bands are split in the same way, i.e., the up-spin state always possesses a higher energy, around the $K$ valley due to the stacking structure and the up-spin states of the lower bands can be found in the Fermi energy.
Thus, the $2R_2$ hetero-trilayer provides spin-selective hole transfer from the NbS$_2$ layer to both top and bottom semiconducting layers.
Moreover, the hetero-trilayer possesses an advantage in comparison with the hetero-bilayer. 
The stacking structure preserves mirror symmetry along the out-of-plane direction and thus two spin states are decoupled in the vdW heterostructure.
Therefore, the $2R_2$ vdW hetero-trilayer enables us to dope spin-valley locked charges into semiconducting TMDC layers with conserving decoupled spin states.

\begin{figure}[htbp]
\begin{center}
 \includegraphics[width=80mm]{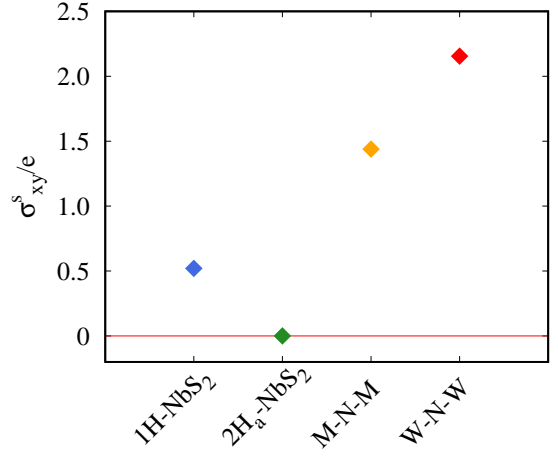}
\caption{The dimensionless spin Hall conductivity of $1H$-NbS$_2$, $2H$-NbS$_2$, $2R_2$ hetero-trilayer MoSe$_2$-NbS$_2$-MoSe$_2$ (M-N-M), and WSe$_2$-NbS$_2$-WSe$_2$ (W-N-W).
 }\label{fig_spin-Hall_conductivity}
\end{center}
\end{figure}
Finally, the spin Hall effect is numerically investigated for the $2R_2$ hetero-trilayers.
In $1H$-TMDCs, electronic states in the $K$ and $K'$ valleys possess large Berry curvatures with the opposite sign to each other.\cite{Xiao2012,Liu2013,Kormanyos2015,Habe2020}
Since the $2R_2$ stacking structure produces spin-selective charge transfer in the two valleys, an enhancement of spin Hall effect is expected in the trilayers. 
The spin Hall conductivity is calculated with using a formula at zero-temperature,\cite{Yao2004} 
\begin{align}
\sigma_{xy}^s=-\frac{e}{\hbar}\int_{BZ}\frac{d^2\boldsymbol{k}}{(2\pi)^2}\sum_{n}\theta(E_F-E_{n\boldsymbol{k}})s^z_n\Omega_n^z(\boldsymbol{k}),
\end{align}
with the step function $\theta(E)$ and the spin eigenvalue $s^z_n=\pm\hbar/2$ of the $n$-th band.
Here, the Berry curvature is represented by
\begin{align}
\Omega_n^z(\boldsymbol{k})=-\sum_{m\neq n}\frac{2\mathrm{Im}\langle\psi_{n\boldsymbol{k}}|v_x|\psi_{m\boldsymbol{k}}\rangle\langle\psi_{m\boldsymbol{k}}|v_y|\psi_{n\boldsymbol{k}}\rangle}{(\omega_{n\boldsymbol{k}}-\omega_{m\boldsymbol{k}})^2},
\end{align} 
with the velocity operator $v_\mu$ and the wave function $|\psi_{m\boldsymbol{k}}\rangle$ with the energy $E_{m\boldsymbol{k}}=\hbar\omega_{m\boldsymbol{k}}$.
The electronic wave functions and operators are obtained with a tight-binding model given by Wannier90.
In this work, five $d$ orbitals in each transition-metal atom and three $p$ orbitals in each chalcogen atom are adopted for spinful Wannier functions.
In Fig.\ \ref{fig_spin-Hall_conductivity}, the numerical calculated conductivity is presented for $1H$-NbS$_2$, $2H_a$-NbS$_2$, $2R_2$ hetero-trilayer MoSe$_2$-NbS$_2$-MoSe$_2$, and WSe$_2$-NbS$_2$-WSe$_2$. 
Although the monolayer NbS$_2$ shows non-zero spin Hall conductivity, $2H_a$-NbS$_2$, the most stable stacking structure, gives no spin Hall effect.
This is because adjacent layers give opposite contributions to spin Hall effect in the conventional stacking structure and inversion symmetry prohibits non-zero spin Hall conductivity.
In the $2R_2$ hetero-trilayers, however, the spin Hall conductivity is three or four times larger than that in monolayer NbS$_2$ without any change of sheet charge density.
The numerical results clearly show the drastic improvement of the electronic spin transport property in the $2R_2$ vdW heterostructure without any extrinsic fabrication.

In conclusion, we theoretically show that the stable vdW heterostructure of metallic and semiconducting atomic layers, NbS$_2$ and MoSe$_2$ (WSe$_2$), is in a novel stacking structure called $2R_2$.
In the $2R_2$ structure, two adjacent layers are stacked in the same way as the $3R$ stacking structure but the period is given by two layers.
The $3R$-like stacking promises the same spin polarization in the $K$ and $K'$ valleys for every layer and the period of two layers provides mirror symmetry along the out-of-plane axis,i.e., spin conservation, for the vdW heterostructures of odd number of layers.
Moreover, the stacking structure induces spin-selective and valley-dependent charge transfer to the semiconducting layers.
Then, we numerically show that the $2R_2$ vdW hetero-trilayers drastically enhance the spin Hall effect of TMDCs though $2H$ stacking, the stable structure of pristine TMDCs, decreases the spin Hall conductivity.

\begin{acknowledgements}
This work was supported by JSPS Grants-in-Aid for Scientific Research No. JP20K05274 and No. JP23K03289.
\end{acknowledgements} 

\bibliography{TMDC}

\end{document}